# FAD-Net: Frequency-Domain Attention-Guided Diffusion Network for Coronary Artery Segmentation using Invasive Coronary Angiography

Nan Mu[a,b,c,#], Ruiqi Song[a,#], Xiaoning Li[a], Zhihui Xu[d], Jingfeng Jiang[e], and Chen Zhao[f,*]

[a]*College of Computer Science, Sichuan Normal University, Chengdu, Sichuan 610101, China.*
[b]*Visual Computing and Virtual Reality Key Laboratory of Sichuan, Sichuan Normal University, Chengdu, Sichuan 610068, China.*
[c]*Education Big Data Collaborative Innovation Center of Sichuan 2011, Chengdu, Sichuan 610101, China.*
[d]*Department of Cardiology, The First Affiliated Hospital of Nanjing Medical, University, Nanjing, China*
[e]*Department of Biomedical Engineering, Michigan Technological University, Houghton, MI 49931, USA.*
[f]*Department of Computer Science, College of Computing and Software Engineering, Kennesaw State University, Marietta, GA 30060, USA.*

## Abstract

**Background:** Coronary artery disease (CAD) remains one of the leading causes of mortality worldwide. However, the accurate identification of arterial stenosis caused by lipid plaques remains a major challenge in early clinical diagnosis and intervention. Precise segmentation of coronary arteries from invasive coronary angiography (ICA) is critical for effective clinical decision-making. Nevertheless, ICA segmentation is challenging due to low contrast, noise interference, and the complex, variable morphology of coronary arteries, resulting in suboptimal performance and impeding accurate stenosis assessment.

**Objective:** This study aims to propose a novel deep learning model based on frequency-domain analysis to enhance the accuracy of coronary artery segmentation and stenosis detection in ICA, thereby offering robust support for the stenosis detection and treatment of CAD.

**Methods:** We propose the Frequency-Domain Attention-Guided Diffusion Network (FAD-Net), which integrates a frequency-domain-based attention mechanism and a cascading diffusion strategy to fully exploit frequency-domain information for improved segmentation accuracy. Specifically, FAD-Net employs a Multi-Level Self-Attention (MLSA) mechanism in the frequency domain, computing the similarity between queries and keys across high- and low-frequency components in ICAs. This enables effective modeling of arterial structures and contextual dependencies, enhancing robustness against background noise. Furthermore, a Low-Frequency Diffusion Module (LFDM) is incorporated to decompose ICAs into low- and high-frequency components via multi-level wavelet transformation. The LFDM progressively denoises and preserves the global vascular topology of ICA images through the diffusion of low-frequency components. Subsequently, it refines fine-grained arterial branches and edges by reintegrating high-frequency details via inverse fusion, enabling continuous enhancement of anatomical precision.

---

[#]N. Mu and R. Song contributed equally to this work.

[*]Corresponding author to provide e-mail: czhao4@kennesaw.edu (C. Zhao).

**Results and Conclusions:** Extensive experiments demonstrate that FAD-Net achieves a mean Dice coefficient of 0.8717 in coronary artery segmentation, outperforming existing state-of-the-art methods. In addition, it attains a true positive rate of 0.6140 and a positive predictive value of 0.6398 in stenosis detection, underscoring its clinical applicability. These findings suggest that FAD-Net holds significant potential to assist in the accurate diagnosis and treatment planning of CAD.

*Keywords:* Frequency-domain attention, Multi-level self-attention, Low-frequency diffusion, Coronary artery disease (CAD), Stenosis detection

## 1. Introduction

Coronary artery disease (CAD), characterized by coronary stenosis resulting from lipid plaque accumulation in the arterial walls, remains one of the leading causes of death worldwide [1]. Stenotic lesions restrict myocardial blood flow, leading to severe clinical events such as angina and myocardial infarction [2, 3]. Currently, percutaneous coronary intervention (PCI) [4] and coronary artery bypass grafting (CABG) [5] are the primary revascularization strategies. Invasive coronary angiography (ICA), regarded as the gold standard for CAD diagnosis, enables precise assessment of coronary anatomy [6]. However, accurate segmentation of coronary vessels in ICA remains a major challenge, as it directly impacts stenosis quantification and clinical decision-making.

Angiographic images often suffer from blurred arterial boundaries due to low contrast and noise, which can lead to unstable segmentation results. Moreover, traditional manual evaluation is inherently subjective and lacks reproducibility. Studies indicate that inter-observer variability in stenosis assessment can reach 20–30% [7]. While rule-based automated methods [8, 9] partially mitigate subjective bias, their generalizability is limited, often resulting in suboptimal segmentation accuracy when applied to complex ICA images. Therefore, enhancing ICA vessel segmentation accuracy is critical to improving the reliability of stenosis detection.

Recent advances in deep learning have significantly advanced automated medical image segmentation. Compared with traditional techniques, deep learning—particularly convolutional neural networks (CNNs [10])—demonstrate superior feature extraction and representation capabilities. These models can automatically learn meaningful patterns from data, achieving remarkable improvements over conventional methods. Nevertheless, most CNN-based models for ICA segmentation (e.g., U-Net [11] and its variants) treat all channel features uniformly and rely on shared weights across spatial positions, limiting their ability to model fine, context-specific features. Such limitations are particularly pronounced in ICA segmentation, where low contrast, noise, and complex vessel morphology pose significant challenges.

The introduction of attention mechanisms [12, 13] offers a promising solution to these limitations. By selectively focusing on salient information and suppressing irrelevant details, attention mechanisms enhance segmentation performance. Typically, they are categorized into spatial attention [14] and channel attention [15]. Spatial attention emphasizes informative spatial regions (e.g., vessel boundaries), while channel attention models inter-channel dependencies to emphasize semantically important features. Despite their benefits, attention-based models still face key limitations in coronary artery segmentation tasks: 1) Sensitivity to background noise, leading to inaccurate identification of target regions. In complex ICA images, existing methods often rely on local feature similarity and lack effective suppression of irrelevant background. This results in misclassification of non-vascular structures, introducing artifacts and boundary ambiguity (see red rectangle in Fig. 1). 2) Insufficient global structural modeling. Conventional spatial domain approaches rely on local similarity computations and struggle to capture long-range dependencies. This leads to fragmented vessel representations, especially in ICA images with complex morphologies and multiple branches (see green arrow in Fig. 1). 3) Weak response to fine details, impairing the detection of capillaries. Although attention mechanisms enhance regional focus, they often fail to capture subtle arterial details such as stenotic areas and branch point textures, thereby reducing fine-grained segmentation precision (see red arrow in Fig. 1).

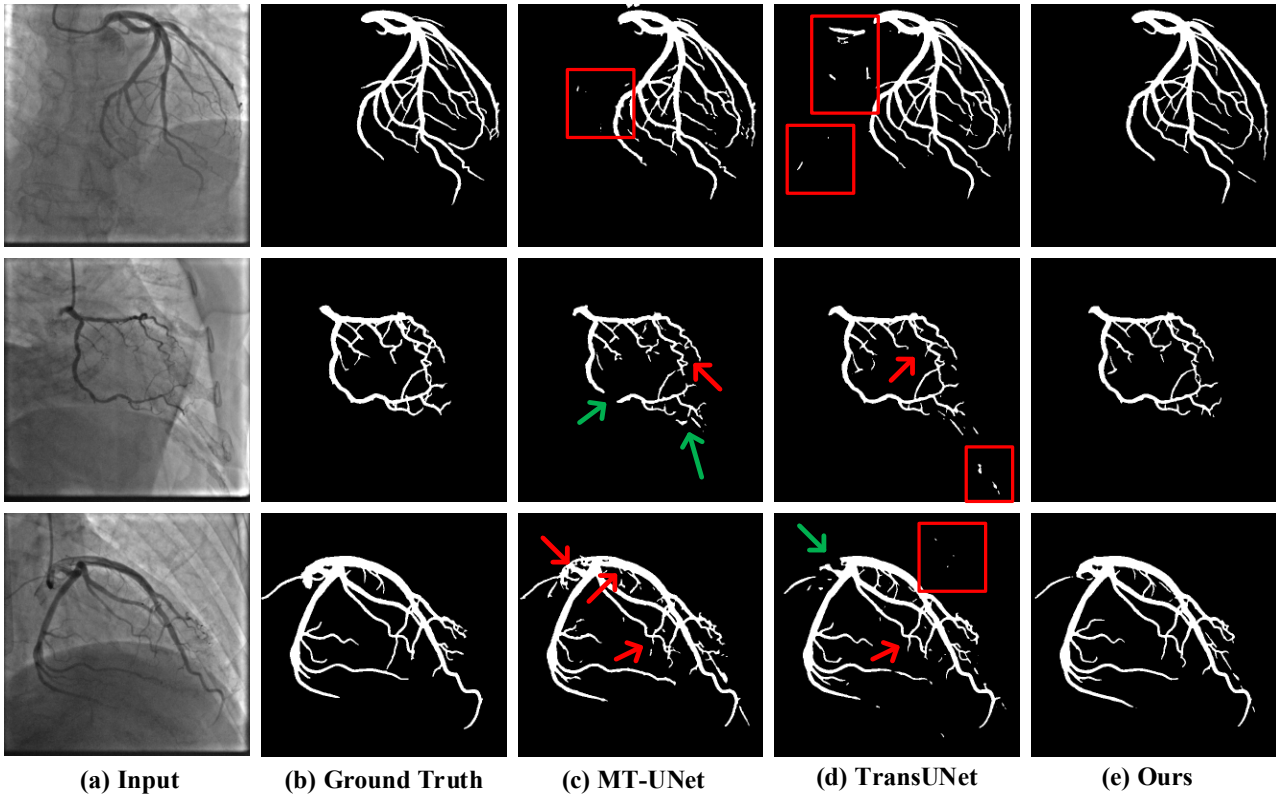

**(a) Input**　　　**(b) Ground Truth**　　　**(c) MT-UNet**　　　**(d) TransUNet**　　　**(e) Ours**

Fig. 1. Visualization of typical failure cases in attention-based models. ICA images and corresponding ground truths are shown in (a) and (b), respectively. Segmentation results of two baseline attention models (MT-UNet [16] and TransUnet [17]) are shown in (c, d), while results from our proposed model are shown in (e).

To address the aforementioned issues, we propose the Frequency-Domain Attention-Guided Diffusion Network (FAD-Net), which integrates frequency-domain processing with deep learning to enhance coronary artery segmentation using ICAs. Specifically, FAD-Net incorporates a Multi-Level Self-Attention (MLSA) module that applies Fourier transforms to the query (Q) and key (K) features. By calculating similarities across different frequency bands, MLSA enables spectrum-driven attention, effectively capturing both local and global contexts. This improves the model's ability to distinguish structural arterial features from background noise and enhances its sensitivity to stable frequency patterns such as edges and textures. Additionally, we explore a Low-Frequency Diffusion Module (LFDM) to balance global structure modeling with local detail enhancement. LFDM decomposes feature maps via multi-level Haar wavelet transforms, and employs learnable scaling factors to progressively enhance low-frequency subbands that encode topological structures of overall arterial tree. Meanwhile, high-frequency subbands (e.g., edges, branch points) are processed via deep convolutional layers to refine fine details, including stenotic features. This dual-focus mechanism allows FAD-Net to collaboratively highlight global vessel topology and preserve intricate vascular features, significantly improving segmentation accuracy and robustness.

Our main contributions are summarized as follows: 1) We propose FAD-Net, a novel deep learning network that integrates frequency-domain attention mechanisms with U-Net for coronary artery segmentation using ICAs. To the best of our knowledge, this is the first approach that jointly applies frequency-domain self-attention and low-frequency diffusion for coronary artery segmentation. 2) We explore frequency-domain modeling to overcome limitations of conventional approaches in low-contrast, noisy, and morphologically complex ICA scenarios. The proposed MLSA module enables effective local-global contextual learning, while LFDM enhances both arterial topology and fine detail preservation—thereby improving the model's robustness and expressiveness. 3) We conduct extensive experiments, including ablation studies, comparative analyses, and stenosis detection evaluations, to validate the effectiveness of FAD-Net. Our model demonstrates superior performance in resisting background interference, highlighting global arterial structures, and accurately extracting fine arterial features. This significantly improves both segmentation accuracy and the positive predictive rate of stenosis detection, offering strong support for clinical CAD diagnosis and treatment. Our code is public available at https://github.com/chenzhao2023/FAD_Net_ICA_BinarySeg.

## 2. Related Work

Current approaches to coronary artery segmentation using ICAs are predominantly based on U-Net architectures, with performance enhancements achieved by integrating various modules. Among these, traditional attention mechanisms and frequency-domain feature processing have emerged as two prominent strategies for improving segmentation quality.

*2.1. Attention Mechanism-Based Medical Image Segmentation*

U-Net has become the foundational architecture for medical image segmentation due to its

encoder–decoder structure and skip connections. However, its limited receptive field and the tendency to lose fine-grained details have motivated the integration of attention mechanisms to enhance performance. Oktay *et al*. [18] introduced a spatial attention gate (AG) into the conventional U-Net to selectively emphasize relevant regions in CT image segmentation. Chen *et al*. [17] proposed TransUNet, which incorporates Transformer blocks into U-Net, combining global context modeling with local feature extraction. Wang *et al*. [16] developed MT-UNet, which leverages a hybrid Transformer module to enhance the network's capability in capturing complex structures. Ji *et al*. [19] introduced DMAGNet, which utilizes a multi-scale normalized channel attention module and skip connections to strengthen channel dependencies and improve segmentation precision.

For vessel segmentation, several models have demonstrated notable improvements. Liu *et al*. [20] proposed AGFA-Net, a deep attention-guided and feature-aggregated network tailored for coronary artery segmentation using coronary CT angiography (CCTA). It refines arterial structures through targeted attention and feature enhancement. Iyer *et al*. [21] designed AngioNet, which improves small arterial branch segmentation via multi-scale feature extraction and a custom-designed attention mechanism. Si *et al*. [22] presented SCSA-CBAM, incorporating collaborative spatial and channel attention to enhance contextual information modeling and segmentation accuracy.

Although these attention-based models have significantly improved the delineation of target regions, they often rely on localized feature interactions and struggle to model long-range dependencies. This limitation becomes particularly apparent in ICA images with low contrast, complex vessel morphology, or small-scale stenotic lesions, where noise and blurred boundaries can lead to misclassification or loss of critical structural information.

*2.2. Frequency-Domain Feature Optimization in Medical Image Segmentation*

Frequency-domain analysis offers a complementary perspective to traditional spatial-domain methods by enabling separation and enhancement of distinct frequency components in medical images. This approach facilitates the targeted modeling of high-frequency features (e.g., stenotic plaques, edges) and low-frequency structures (e.g., vessel trunks, topology), contributing to more robust segmentation. Azad *et al*. [23] proposed FRCU-Net, which integrates discrete cosine transform (DCT) into the decoder stage of U-Net to enhance vascular edge features by emphasizing specific frequency bands. Yan *et al*. [24] introduced a frequency-domain guided attention mechanism that improves multi-scale feature fusion by leveraging frequency-selective modulation. Mu *et al*. [25] developed FACU-Net, a model that constructs a multi-focus attention mechanism in the frequency domain and employs spatially adaptive segmentation strategies to better capture structural nuances. Nam *et al*. [26] proposed MADGNet, which combines multi-frequency and multi-scale attention mechanisms, demonstrating good generalization ability in clinical environments.

These methods demonstrate the potential of frequency-domain representations in overcoming limitations of spatial-domain models, particularly in enhancing contrast, suppressing noise, and retaining fine-grained anatomical details. However, most of the existing approaches either utilize fixed transforms or lack the flexibility to adaptively model frequency components across scales, thereby limiting their effectiveness in complex ICA scenarios with intricate vascular patterns and ambiguous boundaries.

## 3. The Proposed FAD-Net Model

**Overview.** The overall workflow of the proposed FAD-Net is illustrated in Fig. 2. FAD-Net takes an ICA image as input (left of Fig. 2), processes it outputs a segmentation map. This result is then fed into a series of stenosis detection algorithms (right side of Fig. 2), ultimately producing a stenosis comparison map (bottom right of Fig. 2).

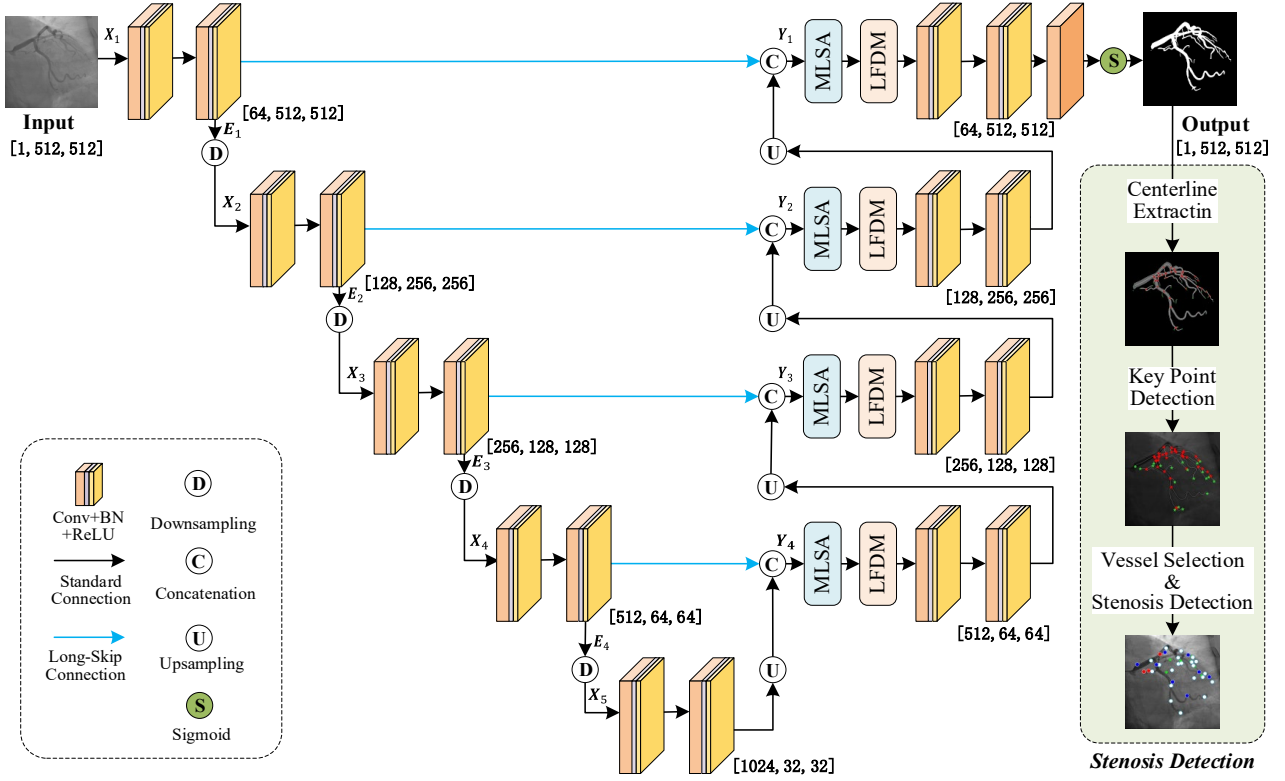

Fig. 2. A schematic overview of the proposed network. The architecture consists of an encoder-decoder backbone with embedded spectral attention modules (i.e., MLSA module and LFDM) designed to improve semantic representation and structural integrity.

*3.1. Encoder-Decoder Structure*

The encoder consists of five hierarchical levels, each comprising a dual 3×3 convolutional block followed by a downsampling operation. In detail, each convolutional block includes two consecutive 3×3 convolutions, each followed by Batch Normalization (BN) [27], Dropout [28] (with a probability of 0.3), and a LeakyReLU activation function [29]. Denoting the input feature map of the $i$-th encoder layer as $X_i \in \mathbb{R}^{C_i \times H_i \times W_i}$, where $i \in \{1,2,3,4,5\}$, $C_i$, $H_i$ and $W_i$ indicate the number of channels, height and width of the feature map. Then, the output of a dual convolutional block is formulated as shown in Eq. (1).

$$E_i = LeakyReLU(BN(Conv_{3\times3}(LeakyReLU(BN(Conv_{3\times3}(X_i))))))). \qquad (1)$$

The downsampling operation is implemented via a 3×3 convolution with a stride of 2, which halves the spatial resolution while preserves channel dimensions, as shown in Eq. (2).

$$X_{i+1} = LeakyReLU(BN(Conv_{3\times3}^{stride=2}(E_i))). \qquad (2)$$

This progressively compresses the spatial features, facilitating the extraction of semantically rich representations for decoding.

The decoder mirrors the encoder in structure, consisting of four upsampling blocks. Let $Y_i \in \mathbb{R}^{C_i \times H_i \times W_i}$ denote the input to the $i$-th decoder layer, which is the aggregated encoded information $E_i$ and the upsampled features $U_{i+1}$ from the deeper decoder layer output $D_{i+1}$. The concatenation operation can be expressed as denoted in Eq. (3).

$$Y_i = Concat(U_{i+1}, E_i), \qquad (3)$$

where $Concat$ represents the concatenation operation.

The upsampled features are obtained through nearest-neighbor interpolation of $D_{i+1}$, followed by a 1×1 convolution to reduce the number of the channels, as shown in Eq. (4).

$$U_{i+1} = Conv_{1 \times 1}(Interpolate(D_{i+1})), \qquad (4)$$

where $Interpolate$ denotes the nearest-neighbor interpolation, which doubles the spatial resolutions.

Each decoder layer begins with two spectral enhancement modules: the Multi-Level Self-Attention (MLSA) module and the Low-Frequency Diffusion Module (LFDM). Both modules are sequentially applied to model global context dependencies and enhance low-frequency structural information propagation. After spectral enhancement, the refined features are further processed by another dual convolutional block, which includes a 3×3 convolution, followed by Batch Normalization and a LeakyReLU activation function, thus enabling the extraction of deeper semantic features and improving the non-linear feature representation capability.

In the context of the coronary artery segmentation, this architecture effectively restores spatial details and enhances the clarity of segmentation boundaries, improving the overall segmentation performance.

*3.2. Multi-Level Self-Attention (MLSA) Module*

Unlike traditional spatial-domain attention-based mechanisms that rely on dot-product correlations between query and key matrices, MLSA introduces a frequency-domain attention framework inspired by the convolution theorem, which states that spatial-domain convolution corresponds to pointwise multiplication in the frequency domain. This paradigm enables both efficient and structured modeling of global and local dependencies. Specifically, MLSA performs pointwise multiplications between the query (Q) and key (K) features across different frequency bands—namely, low- and high-frequency domains—to adaptively enhance feature saliency at both global and local levels. After multi-band decomposition of Q and K, the pointwise multiplication attention in the low-frequency subbands tends to assign higher weights to trunk regions in ICA images, thereby reinforcing attention to the overall anatomical structure and preserving vascular topology. In contrast, attention in the high-frequency subbands highlights salient edges and fine details, improving the model's ability to distinguish vessel branches and textures.

Overall, MLSA decomposes Q and K into low- and high-frequency subbands and computes attention weights separately within each, effectively modeling both global and local dependencies. This dual-band strategy enables the model to balance global structural consistency with fine-grained detail representation.

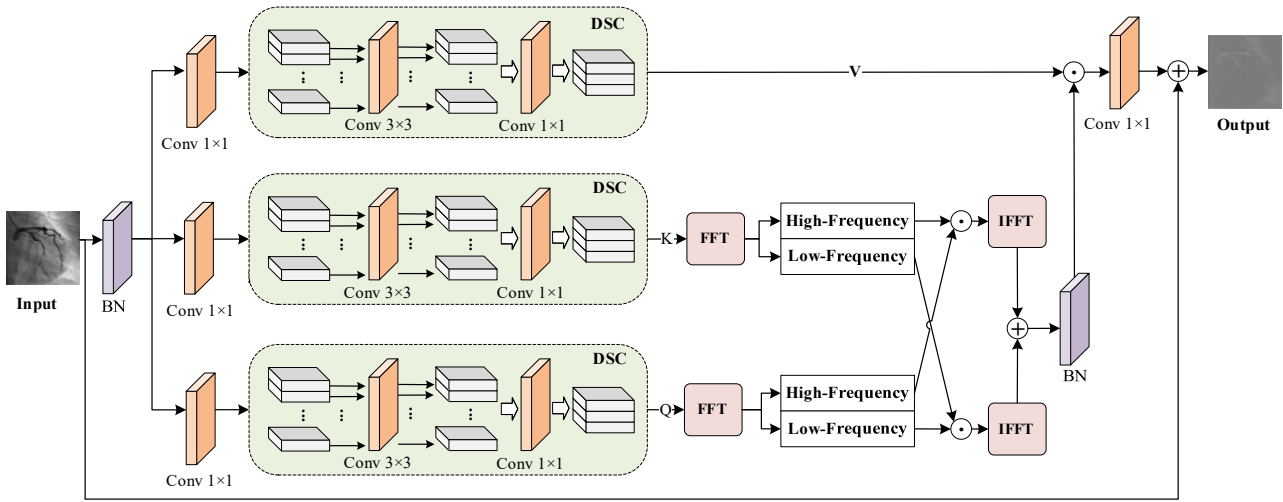

Fig. 3. The structure of the proposed MLSA module.

As depicted in Fig. 3, the MLSA module (a Frequency-Split Attention Structure) comprises the following core components: The query ($F_q$), key ($F_k$), and value ($F_v$) features are extracted via 1×1 convolution and 3×3 depthwise separable convolution (DSC). Subsequently, the query and key features are transformed into the frequency domain using Fast Fourier Transform (FFT). Based on the radial distance of each frequency component from the center, the spectrum is divided into low- and high-frequency subbands using a normalized radius threshold of 0.5. Specifically, components with a radius less than 0.5 are classified as the low-frequency subband, while those with a radius greater than 0.5 belong to the high-frequency subband. Subband-specific spatial attention maps (denoted as $F_{low}$ and $F_{high}$) are computed via element-wise multiplication in the frequency domain, followed by inverse FFT (IFFT):

$$F_{low} = \mathcal{F}^{-1}\big(\mathcal{F}(F_q)_{low} \odot \mathcal{F}(F_k)_{low}\big), \tag{5}$$

$$F_{high} = \mathcal{F}^{-1}\big(\mathcal{F}(F_q)_{high} \odot \mathcal{F}(F_k)_{high}\big), \tag{6}$$

$$F_{ATT} = F_{low} \oplus F_{high}, \tag{7}$$

where $\mathcal{F}(\cdot)$ and $\mathcal{F}^{-1}(\cdot)$ denote the FFT and IFFT operations, respectively. $\odot$ and $\oplus$ represent element-wise multiplication and summation, respectively.

This approach exploits the distinct advantages of different frequency bands: low-frequency bands maintain global structure continuity, while high-frequency bands emphasize fine-grained details and textures [30]. Compared with conventional spatial-domain attention mechanisms [31], which typically compute attention weights via the dot-product correlation between the query and key matrices with a computational complexity of $O(N^2)$, MLSA significantly reduces the computational burden to $O(N\log N)$ by transforming the dot-product operation into an element-wise multiplication in the frequency domain. Moreover, by decomposing the query and key features into low- and high-frequency subbands, MLSA decouples global semantic information (i.e., structural context) from fine-grained details such as edges, textures, and noise. This separation enables noise-free attention computation in the low-frequency subband, thereby maximizing the integrity of the primary vascular trunk. Meanwhile, attention in the high-frequency subband enhances edge and texture representations while suppressing irrelevant noise. This frequency-aware self-attention mechanism facilitates precise target discrimination while effectively mitigating background interference.

Notably, MLSA departs from the traditional attention mechanism's reliance on the Softmax operation for computing attention weights—typically formulated as $w = \text{Softmax}(QK^T) \times V$. While Softmax normalizes attention scores into a probability distribution, it often induces sparsity across spatial or channel dimensions, potentially suppressing low-response yet semantically important structural features. In contrast, MLSA computes attention weights directly in the frequency domain using the formulation $w = \text{IFFT}(\text{FFT}(Q) \odot \text{FFT}(K)) \times V$, where $\odot$ denotes element-wise multiplication. This approach inherently captures structural correlations between spatial patches and implicitly constructs attention maps without imposing normalization constraints. As a result, MLSA preserves richer and more fine-grained feature responses, enhancing the model's capacity to recognize subtle yet critical anatomical patterns.

The final attention-weighted output is generated by multiplying the frequency attention map $F_A$ with the value feature $F_v$, followed by a 1×1 convolution and residual addition with the decoder input $Y_i$:

$$V_{ATT} = \mathcal{L}(F_{ATT}) \cdot F_v, \tag{8}$$

$$Y_i^{MLSA} = Y_i + Conv_{1\times1}(V_{ATT}). \tag{9}$$

where $\mathcal{L}(\cdot)$ represents a linear transformation applied to the attention map.

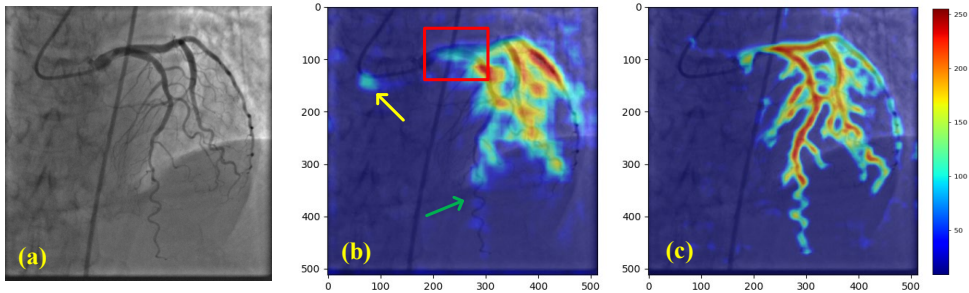

Fig. 4. The visual evidence of the MLSA module's effectiveness. (a) shows the input image. (b) and (c) are Grad-CAM heatmaps [32] generated by standard convolution and MLSA, respectively. Red regions indicate strong activations critical for decision-making; yellow and green indicate moderate importance; blue denotes minimal contribution.

As shown in Fig. 4, MLSA enhances segmentation performance and robustness, particularly in conronary artery segmentation using ICAs. Its frequency-decomposed, hierarchical modeling framework effectively enhances key arterial structures while suppressing irrelevant background noise and imaging artifacts. Specifically, the low-frequency branch captures the global arterial topology, ensuring structural continuity and reducing discontinuity artifacts (as highlighted by the red box in Fig. 4(b)). The high-frequency branch enhances fine structural details such as capillaries (see green arrow in Fig. 4) while mitigating the influence of background and noise (see yellow arrow in Fig. 4).

This "frequency decomposition + hierarchical modeling" strategy not only preserves the anatomical structure of vessels but also significantly boosts segmentation precision and robustness. In the FAD-Net architecture, the MLSA module is embedded after the skip connections in the decoder stage, improving feature fusion during upsampling. By capturing attention across different spectral bands (i.e., low- and high-frequency domains), MLSA delivers enhanced sensitivity to both vessel trunks and branches while maintaining resilience against irrelevant signals. As demonstrated in Fig. 4(c), the proposed module yields sharper, more continuous vascular representations than conventional convolution-based approaches.

*3.3. Low-Frequency Diffusion Module (LFDM)*

The Low-Frequency Diffusion Module (LFDM) is a feature enhancement module based on a cascaded architecture of Wavelet Transform (WT) and Inverse Wavelet Transform (IWT) [30], as illustrated in Fig. 5. Unlike the MLSA module, which focuses on the self-attention among different spectral bands, LFDM emphasizes the progressive diffusion of low-frequency features to capture and preserve the global topological structure of the vascular tree. Meanwhile, it enhances high-frequency components—such as arterial edges, fine branches, and stenotic lesions—through cascaded depthwise separable convolutions, achieving a fine-grained balance between structural integrity and detail representation. Specifically, LFDM employs multi-stage wavelet transforms to incrementally extract increasingly lower-frequency components, enabling the preservation of the overall vascular topology. For high-frequency components, the module applies cascaded depthwise separable convolutions (DSC [33], as illustrated in Fig. 5), where depthwise convolutions independently extract spatial features (e.g., edges and textures) within each channel without interference from others. This mechanism effectively captures critical local details in ICA images—such as vessel boundaries, subtle branches, and stenotic regions—thus reinforcing precise modeling of local anatomical structures.

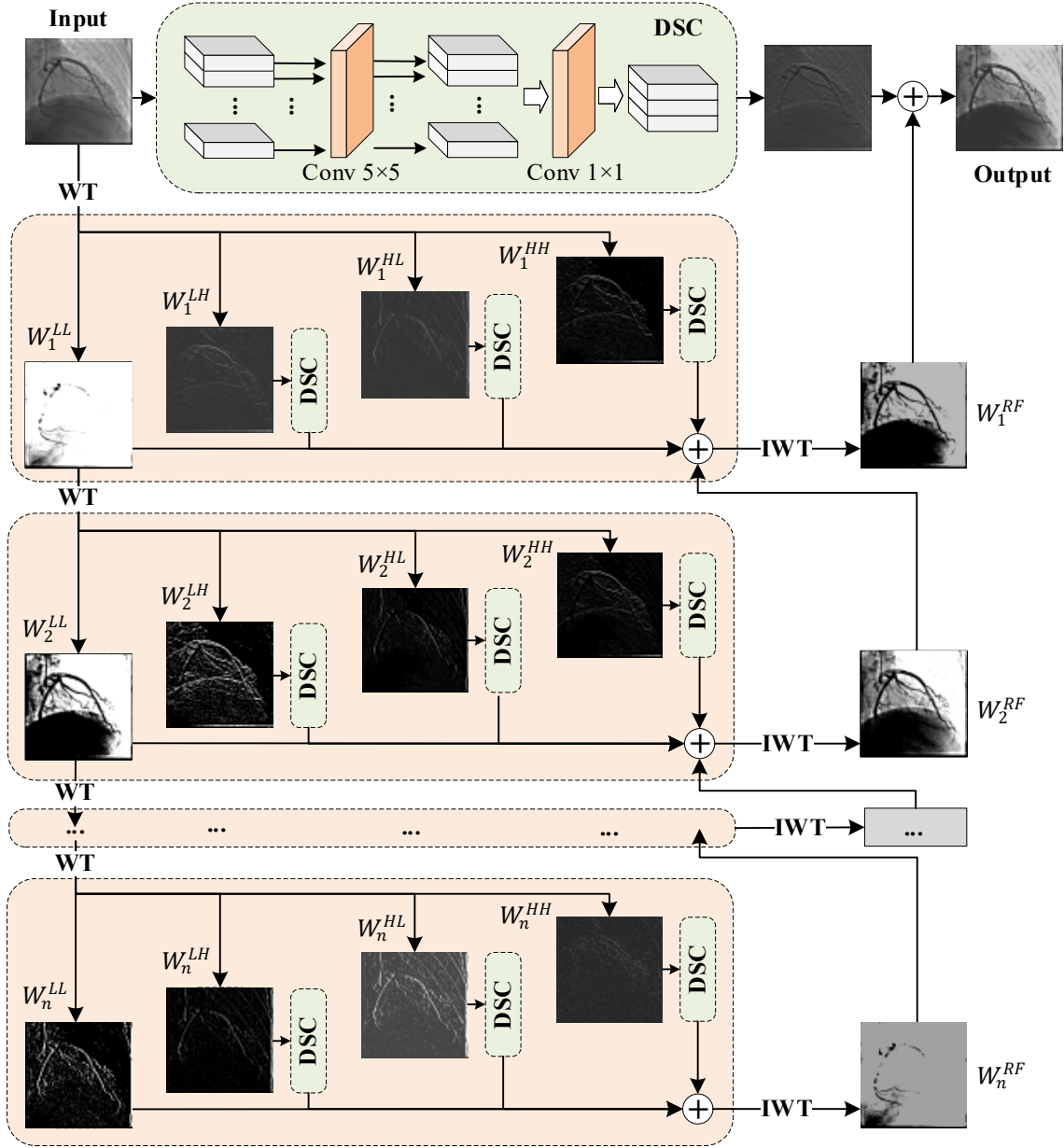

Fig. 5. The structure of the proposed Low-Frequency Diffusion Module (LFDM).

The LFDM employs the Haar wavelet to perform frequency decomposition, generating one low-pass filter $f_{LL} = \frac{1}{\sqrt{2}}[1\ \ 1]$ and three high-pass filters: $f_{LH} = \frac{1}{\sqrt{2}}[1\ -1]$, $f_{HL} = \frac{1}{\sqrt{2}}\begin{bmatrix} 1 & 1 \\ -1 & -1 \end{bmatrix}$, and $f_{HH} = \frac{1}{\sqrt{2}}\begin{bmatrix} 1 & -1 \\ -1 & 1 \end{bmatrix}$. The low-pass filter $f_{LL}$ captures the approximate (low-frequency) components of the signal, preserving the overall anatomical structure and smooth intensity transitions. In contrast, the high-pass filters extract directional high-frequency information: $f_{LH}$ emphasizes horizontal edges, $f_{HL}$ enhances vertical contours, and $f_{HH}$ highlights diagonal structures. This decomposition enables the model to better delineate vessel boundaries and capture subtle textures critical for fine-grained vascular analysis. These filters generate four sub-bands $W_j^{LL}, W_j^{LH}, W_j^{HL}, W_j^{HH}$ at layer $j \in \{1, \ldots, n\}$, where $W_j^{LL}$ is the low-frequency trunk component preserving global structure, $W_j^{LH}, W_j^{HL}$, and $W_j^{HH}$ represent the high-frequency components capturing horizontal, vertical, and diagonal textures such as edges, branches, and local texture details. To extract progressively global and multi-scale semantic information, LFDM performs recursive decomposition and diffusion on $W_j^{LL}$, forming a low-frequency diffusion pyramid:

$$W_j^{LL}, W_j^{LH}, W_j^{HL}, W_j^{HH} = \begin{cases} \mathcal{W}(Y_i^{MLSA}), & if\ j = 1, \\ \mathcal{W}(W_{j+1}^{LL}), & if\ j = 2, \ldots, n, \end{cases} \quad (10)$$

where $\mathcal{W}(\cdot)$ denotes the Wavelet Transform (i.e., Haar wavelet). By repeatedly applying Wavelet Transforms to the low-frequency components, LFDM progressively decomposes them into even lower-frequency parts along with their corresponding high-frequency details, thereby achieving gradual low-frequency diffusion. At each decomposition level, the low-frequency output is fed into the next stage of decomposition, while the high-frequency components are refined using depthwise separable convolutions (DSC). During the forward diffusion of low-frequency signals, noise and fine details are gradually suppressed, allowing the preservation of the vessel tree's global topological structure. In the subsequent reverse fusion stage, fine-grained features of the ICA are progressively reconstructed by integrating detail information extracted from the high-frequency branches at each level.

The high-frequency subbands $W_j^{LH}, W_j^{HL}$, and $W_j^{HH}$ at each layer $j$ are refined and denoised by applying a 5×5 depthwise separable convolution (stride=2) operator. These refined sub-bands, together with the low-frequency trunk $W_j^{LL}$ and the reconstructed features from the next layer $W_{j+1}^{RF}$, are aggregated and passed through the Inverse Wavelet Transform $\mathcal{W}^{-1}(\cdot)$ to yield the reconstructed feature at layer $j$:

$$W_j^{RF} = \mathcal{W}^{-1}(W_j^{LL} + \mathcal{D}(W_j^{LH}) + \mathcal{D}(W_j^{HL}) + \mathcal{D}(W_j^{HH}) + W_{j+1}^{RF}), \quad (11)$$

where $\mathcal{D}(\cdot)$ denotes depthwise separable convolution for preserving spatial detail and extracting multi-scale representations. Thanks to the additive linearity of wavelet transforms, low-frequency features are progressively fused in a directionally consistent and lossless manner, enabling topology-preserving vessel reconstruction with enhanced structural completeness and detail clarity.

The final output of LFDM is obtained by fusing the reconstructed features with the attention features from the MLSA module:

$$Y_i^{LFDM} = \mathcal{D}(Y_i^{MLSA}) \oplus W_1^{RF}, \tag{12}$$

where $\oplus$ represent element-wise summation.

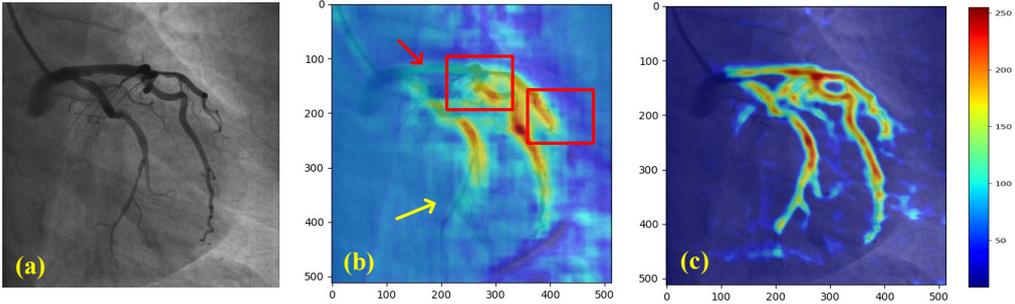

Fig. 6. A comparative heatmap analysis of LFDM. (a) is the original image. (b) and (c) are heatmaps generated by standard convolution and LFDM, respectively.

The response heatmap of the LFDM-processed images is illustrated in Fig. 6. As depicted, the LFDM is positioned after the MLSA module and before the upsampling stage (refer to Fig. 2), serving as a critical component in the decoder. Distinct from conventional decoding strategies that simply fuse high-resolution features, LFDM introduces a novel mechanism centered on multi-level diffusion and reconstruction of low-frequency information to enhance global structural perception and detailed representation. Initially, LFDM applies wavelet decomposition to the input features, isolating the low-frequency backbone component $W_j^{LL}$, which encapsulates the global arterial topology. This component undergoes progressive diffusion across multiple levels, effectively propagating and emphasizing the structural context (highlighted by the red arrow in Fig. 6(b)). During reconstruction, these low-frequency features are iteratively fused from top to bottom at each layer, enabling cross-scale structural recovery and seamless semantic continuity.

Concurrently, high-frequency subbands—such as those representing edges, fine branches, and stenotic regions (yellow arrow in Fig. 6(b))—are refined through depthwise separable convolutions, facilitating precise local detail modeling. In intracoronary angiography (ICA) images, essential structural and connectivity cues are predominantly encoded in the low-frequency spectrum, while fine details and noise reside in the high-frequency components. By amplifying critical low-frequency signals and selectively enhancing informative high-frequency features while suppressing noise, LFDM significantly boosts segmentation robustness and precision. This dual strategy of "low-frequency diffusion + high-frequency refinement" allows LFDM to mitigate typical limitations of traditional fusion methods, such as information redundancy and artifact amplification, ultimately supporting more accurate and coherent feature reconstruction.

As evidenced in Fig. 6, the LFDM-generated heatmap demonstrates a pronounced activation in the low-frequency channels aligned with the vascular backbone, while maintaining stable and focused responses in high-frequency channels associated with capillaries. Compared to standard convolutions, LFDM markedly reduces background activation (as shown by the red rectangle in Fig. 6(b)), underscoring its superiority in structural guidance and interference suppression.

*3.4. Loss Function Design*

To address the topological continuity and noise robustness required for ICA vessel segmentation, the Mean Squared Error (MSE) is employed as the primary loss function:

$$L_{MSE} = \frac{1}{N}\sum_{i=1}^{N}(y_i - \hat{y}_i)^2, \tag{13}$$

where $N$ is the number of pixels, $y_i$ is the ground truth, and $\hat{y}_i$ is the predicted value.. The MSE loss function offers smooth and continuous gradients that promote spatial consistency within local regions of the prediction. Moreover, its strong penalization of large deviations makes it effective in mitigating noise and enhancing overall prediction stability. During training, gradients derived from this loss are used to iteratively optimize model parameters through backpropagation.

## 4. Experiments

This section describes the experimental setup, performance analysis (including comparative experiments and ablation studies), and stenosis detection. The results validate the accuracy and robustness of our method.

*4.1. Experimental Setup*

**Training Details:** The FAD-Net model was implemented in the PyTorch framework and trained on dual RTX 3090 GPUs with a combined 48 GB of VRAM. We used the Adam optimizer [34] with an initial learning rate of $1 \times 10^{-4}$, $\beta_1 = 0.9$, $\beta_1 = 0.999$, and a weight decay of $1 \times 10^{-5}$. A multi-stage learning rate scheduling strategy using StepLR was adopted: the learning rate remained at $1 \times 10^{-4}$ for the first 10 epochs to focus on major vessel learning. From epochs 10 to 30, it was halved every 10 epochs to refine predictions for secondary branches. After 30 epochs, the rate was reduced to $3.125 \times 10^{-5}$ to emphasize microvessel edge refinement and structural detail preservation, thereby enhancing segmentation precision and model stability.

**Image Dataset:** The dataset consists of ICA images from 99 patients who underwent examinations at Jiangsu Provincial Hospital, China, between February 26 and July 18, 2019. Images were acquired using a Siemens AXIOM-Artis interventional angiography system at 15 frames per second. Each image had a resolution of 512×512 pixels, with a pixel spacing between 0.258 mm and 0.390 mm. In total, we collected 187 left coronary artery (LCA) images and 127 right coronary artery (RCA) images, each manually annotated by experienced interventional cardiologists. Up to five standard projection views were selected per patient, with one frame per view used for segmentation. Table 1 summarizes the distribution of ICA views and image counts.

Table 1: Distribution of ICA views and image counts. (LCA: Left Coronary Artery; RCA: Right Coronary Artery; LAO: Left Anterior Oblique; RAO: Right Anterior Oblique; CRA: Cranial; CAU: Caudal)

| View | LAO+CAU | LAO+CRA | RAO+CAU | RAO+CRA | Total |
| --- | --- | --- | --- | --- | --- |
| LCA | 82 | 82 | 119 | 120 | 403 |
| RCA | 80 | 83 | 23 | 27 | 213 |

Evaluation Metrics: We used five voxel-level classification metrics to assess egmentation performance: Dice Similarity Coefficient (DICE), Sensitivity, Specificity, 95% Hausdorff Distance (HD95), and Average Symmetric Surface Distance (ASSD). These metrics collectively measure both volumetric overlap and surface agreement between the predicted and ground truth masks. All metrics were computed on a per-slice basis and averaged over the test set. The formulas for these metrics are defined as:

$$DICE = \frac{2TP}{2TP+FP+FN}, \tag{14}$$

$$Sensitivity = \frac{TP}{TP+FN}, \tag{15}$$

$$Specificity = \frac{TN}{TN+FP}, \tag{16}$$

$$HD95 = max\{P_{95}(min_{b \in B} \parallel a - b \parallel), P_{95}(min_{b \in B} \parallel a - b \parallel)\}, \tag{17}$$

$$ASSD = \frac{1}{|A|+|B|}(\sum_{a \in A} min_{b \in B} \parallel a - b \parallel + \sum_{b \in B} min_{a \in A} \parallel b - a \parallel), \tag{18}$$

where TP, FP, TN, and FN represent true positive, false positive, true negative, and false negative rates, respectively. $P_{95}$ represent the 95th percentile of the distance set, and $A$, $B$ are the surface point sets of the predicted and ground truth segmentations. $\parallel \cdot \parallel$ represents the Euclidean distance.

*4.2. Segmentation Performance Evaluation*

We compared the proposed FAD-Net with four state-of-the-art deep learning segmentation models: Swim-Unet [35], U-Net [11], U-Net++ [36], and DSCNet [37]. The comparison was conducted using the five evaluation metrics mentioned above.

**Quantitative Evaluation.** Table 2 presents the numerical results across the five metrics. FAD-Net achieved the best performance overall, particularly excelling in surface-based metrics. It recorded the lowest HD95 (8.04 mm) and ASSD (1.46 mm), indicating superior accuracy in vessel boundary delineation and topological continuity. Its Dice score (87.17%) reflects high overlap with the ground truth, and its strong sensitivity (86.34%) and specificity (99.49%) illustrate a well-balanced capability in detecting small vessels while suppressing false positives. Compared with other models, DSCNet and U-Net++ performed moderately well but struggled with microvascular continuity. Swim-Unet underperformed across all metrics. In summary, FAD-Net demonstrates outstanding performance in both macro- and micro-vascular segmentation, highlighting its capability in modeling intricate anatomical structures.

Table 2: Quantitative comparison of different segmentation models. (↑ means higher is better; ↓ means lower is better)

| Model | DICE (%) ↑ | Sensitivity (%) ↑ | Specificity (%) ↑ | HD95 (pixel) ↓ | ASSD (pixel) ↓ |
|---|---|---|---|---|---|
| Swim-Unet | 72.07±2.11 | 73.00±3.02 | 98.35±0.41 | 21.3434±2.18 | 3.3651±0.42 |
| U-Net | 85.03±1.46 | 83.49±2.21 | 99.38±0.26 | 11.6230±1.23 | 2.0581±0.31 |
| U-Net++ | 85.57±1.29 | 83.53±2.15 | 99.41±0.24 | 9.3624±1.08 | 1.9292±0.28 |
| DSCNet | 85.06±1.32 | **91.84±2.09** | 99.12±0.34 | 11.9087±1.35 | 2.0040±0.33 |
| FAD-Net | **87.17±1.25** | 86.34±2.03 | **99.49±0.21** | **8.0354±0.97** | **1.4557±0.24** |

**Qualitative Evaluation.** Fig. 7 presents visual comparisons of segmentation results from different models on both LCA (top two rows) and RCA (bottom two rows) views. The yellow regions indicate the predicted segmentations. As illustrated, FAD-Net shows a clear advantage in preserving complex bifurcations, tortuous artery paths, and fine arterial structures (highlighted by blue arrows). In areas where other models break continuity in small branches or misclassify background, FAD-Net maintains the anatomical coherence and produces smoother boundaries. Compared to U-Net++ and DSCNet, FAD-Net yields cleaner segmentations, especially in bifurcation zones and low-contrast regions. These qualitative results align with the superior performance metrics reported in Table 2.

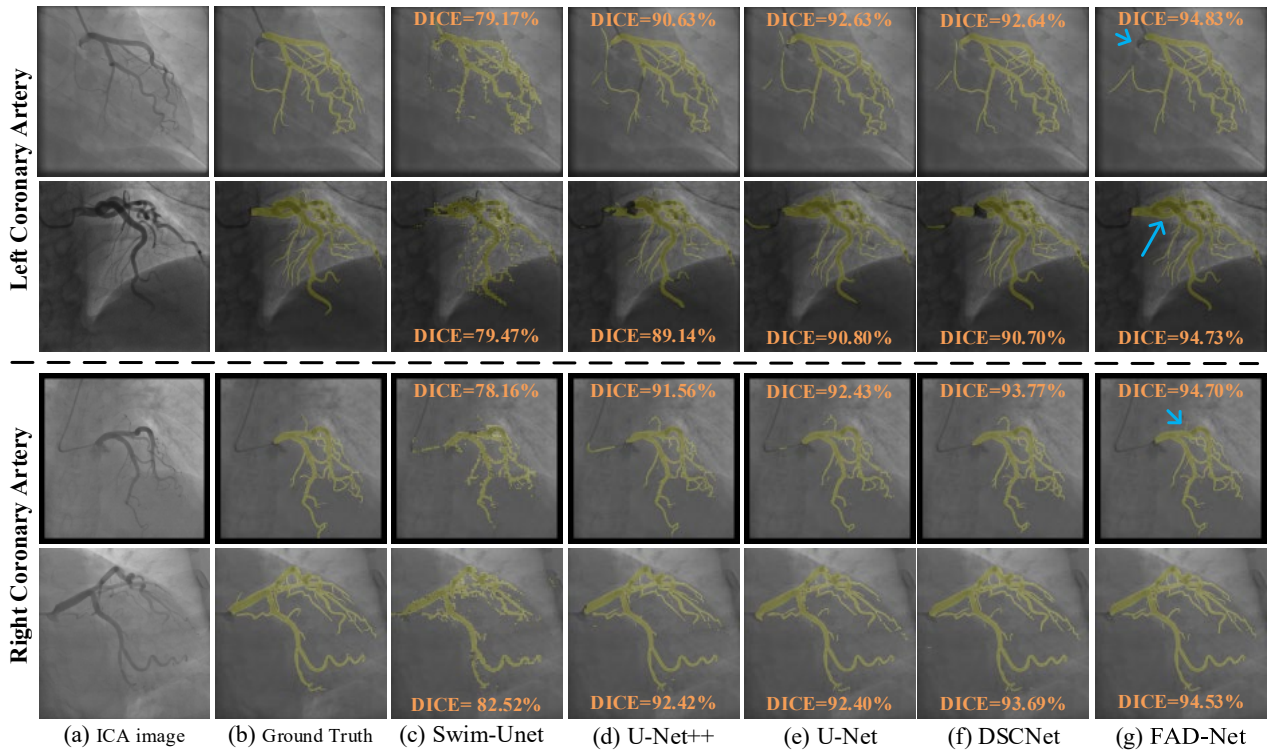

Fig. 7. Visual comparison of segmentation results from different models for left and right coronary arteries. Subfigure (a) shows ICA images, (b) shows ground truth annotations. (c-g) are the segmentation results, with DICE metrics annotated.

### 4.3. Ablation Studies

To validate the architectural advantages of FAD-Net, we conducted a series of ablation experiments based on four model variants: Plain U-Net (U); U-Net with only MLSA (U+MLSA); U-Net with only LFDM (U+LFDM); FAD-Net, integrating both MLSA and LFDM modules.

**Quantitative Results.** Table 3 summarizes the performance of each variant across five evaluation metrics, clearly highlighting the contributions of the MLSA and LFDM modules to segmentation accuracy. Compared to the baseline U-Net, incorporating the MLSA module enhances the Dice score from 85.03% to 86.23% and reduces ASSD from 2.06 mm to 1.64 mm, reflecting improved capability in capturing fine structural details. Adding the LFDM module yields further performance gains, notably reducing HD95 from 11.62 mm to 7.53 mm and ASSD to 1.50 mm, suggesting enhanced modeling of vessel boundaries and spatial continuity. When both modules are integrated into FAD-Net, the model achieves optimal results across most metrics, particularly excelling in DICE (87.17%) and ASSD (1.46 mm).

Table 3 : Quantitative results of different ablation variants. (↑ means higher is better; ↓ means lower is better)

| Models | DICE (%) ↑ | Sensitivity (%) ↑ | Specificity (%) ↑ | HD95 (pixel) ↓ | ASSD (pixel) ↓ |
|---|---|---|---|---|---|
| U | 85.03±1.48 | 83.49±2.74 | 99.38±0.19 | 11.6230±1.67 | 2.0581±0.28 |
| U+MLSA | 86.23±1.31 | 86.33±2.58 | 99.36±0.26 | 9.0452±1.44 | 1.6429±0.25 |
| U+LFDM | 86.82±1.27 | 86.06±2.59 | 99.46±0.28 | 7.5311±1.03 | 1.5011±0.21 |
| FAD-Net | **87.17±1.25** | **86.34±2.03** | **99.49±0.21** | **8.0354±0.97** | **1.4557±0.24** |

**Qualitative Results.** Fig. 8 illustrates the segmentation results of each variant for both the left and right coronary arteries. Among the models, FAD-Net (Fig. 8(f)) achieves the smoothest and most continuous vessel boundaries, especially in small branches and bifurcations, while substantially reducing segmentation artifacts. In contrast, the other variants (Fig. 8(c)-(e)) show varying degrees of disconnections and false positives, highlighted in red rectangles. The MLSA module enhances robustness to background noise and preserves the global vessel structure. The LFDM module improves detection of fine edges and smaller vessels. Their combination in FAD-Net enhances both main vessel segmentation and fine branch continuity. Notably, in the regions indicated by blue arrows (e.g., curved or thin branches), FAD-Net more accurately segments vessels that other variants miss or blur.

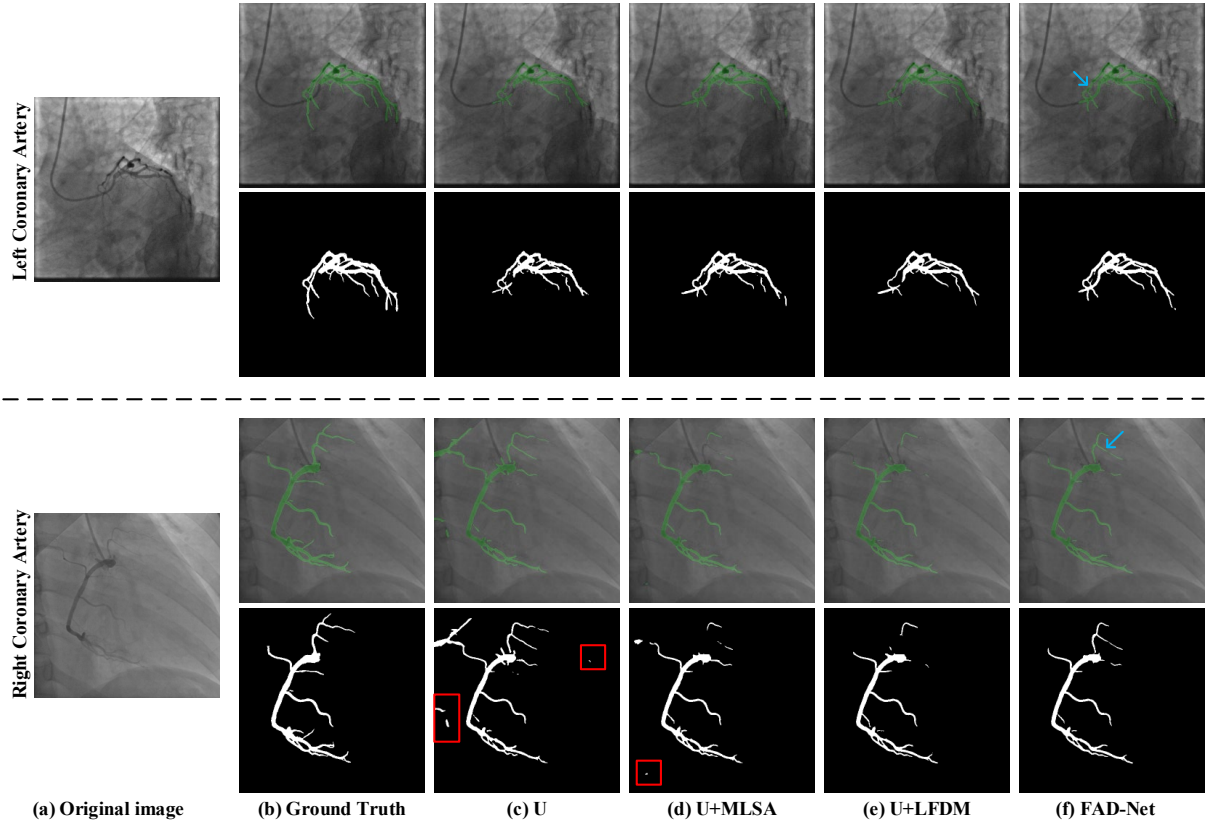

Fig. 8. Visual comparison of ablation results on left and right coronary artery segmentation. (a) and (b) display the original images and the ground truths. (c-f) represent the outputs from U, U+MLSA, U+LFDM, and FAD-Net, respectively. Each case includes two rows of visualizations: the top row overlays model predictions (green) on the original angiograms, while the bottom row shows the corresponding binary vessel maps.

**Attention Visualization.** Fig. 9 provides heatmaps of attention regions generated by different models on the angiographic images. The red areas denote strong activations crucial to decision-making. Compared to other variants, FAD-Net's attention maps cover a broader and more accurate vessel region, capturing clear vessel contours. This confirms the model's improved focus on semantically meaningful image components. Together, Fig. 8 and Fig. 9 substantiate the effectiveness of combining multi-level frequency-domain attention with low-frequency diffusion, which aligns with the quantitative improvements shown in Table 3.

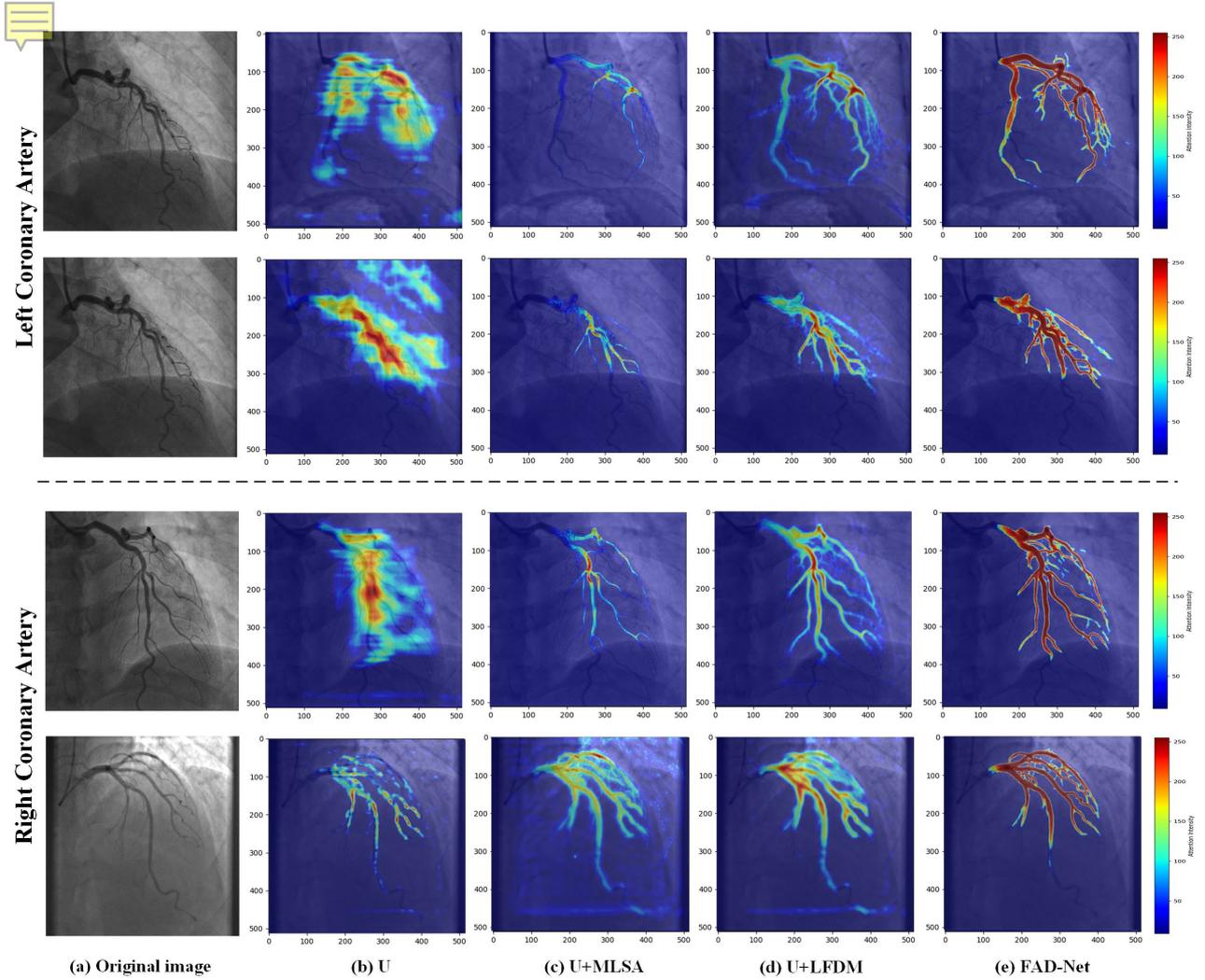

(a) Original image  (b) U  (c) U+MLSA  (d) U+LFDM  (e) FAD-Net

Fig. 9. Attention heatmaps highlighting the regions focused on by different ablation variants during segmentation. (a) Original image; (b-e) show attention maps from each variant. Red: high-activation regions; Yellow/green: moderate; Blue: low-activation areas.

*4.4. Stenosis Detection and Evaluation*

To evaluate the clinical applicability of segmentation results, we introduce a stenosis detection algorithm [38] that quantitatively assesses the model's capability in identifying pathological vascular narrowing. This process aims to bridge the gap between anatomical segmentation and diagnostic utility by validating whether the automatically segmented contours can effectively localize and characterize arterial stenosis. The algorithm follows a centerline-based analysis to extract vessel topology, measure luminal diameter variation, and detect significant constriction. The algorithm for detecting coronary artery stenosis unfolds in the following steps:

| **Algorithm:** Artery Stenosis Detection and Evaluation |
|---|
| **Input:** Segmented binary artery image $I$, Detection radius $r$, Centerline length threshold $L_{thresh} = 20$, Diameter threshold $D_{thresh} = 1.8$ mm, Stenosis ratio threshold $b_{thresh} = 0.1$ |
| **Output:** Stenosis severity per arterial segment, Evaluation metrics (TPR, PPV, ARMSE, and RRMSE) |
| **1. ExtractCenterline** ($I$)<br>    Apply morphological thinning until skeleton stabilizes<br>    Return single-pixel-width centerline $C$ |
| **2. ComputeDiameters** ($I$, $C$)<br>    Perform Euclidean Distance Transform: EDT ← EuclideanDistanceTransform ($I$)<br>    **For** each point $p \in C$:<br>        $d(p) \leftarrow 2 \times \text{EDT}(p)$<br>    **End**<br>    **Return** diameter list $d$ |
| **3. DecomposeSegments** ($C$)<br>    Label node degrees: endpoint (1), connector (2), bifurcation (>2)<br>    Decompose centerline via edge linking to generate segments<br>    **Return** segment list |
| **4. DetectStenosis** (segments, $d$)<br>    **For** each segment $S$:<br>        **If** max($d \in S$) < $D_{thresh}$ or length($S$) < $L_{thresh}$: skip<br>        $d_{min} \leftarrow$ min(local minima of d($S$))<br>        $d_{ref} \leftarrow$ max(local maxima of d($S$)) (or global max if none)<br>        $b \leftarrow (1 - d_{min}/d_{ref}) \times 100\%$<br>        **If** $b \geq b_{thresh}$: mark as stenotic with severity $b$<br>    **End**<br>    **Return** marked stenotic segments |
| **5. MatchWithGroundTruth** (stenotic segments)<br>    **For** each predicted segment $S_{pred}$:<br>        **If** endpoint distances match any ground truth (GT) segment within $r$: mark as matched<br>        **Else if** nearest GT stenotic point within $r$: assign correspondence<br>    **End**<br>    **Return** match results |
| **Main Routine**<br>    $C \leftarrow$ ExtractCenterline ($I$)<br>    $d \leftarrow$ ComputeDiameters ($I$, $C$)<br>    segments ← DecomposeSegments ($C$)<br>    stenotic segments ← DetectStenosis (segments, $d$)<br>    MatchWithGroundTruth (stenotic segments)<br>    **Return** stenotic segments and evaluation metrics |

To quantify detection performance, we adopt four standard metrics: True Positive Rate (TPR), Positive Predictive Value (PPV), Absolute Root Mean Square Error (ARMSE), and Relative Root Mean Square Error (RRMSE). These are computed as:

$$TPR = \frac{TP}{TP+FN}, \tag{19}$$

$$PPV = \frac{TP}{TP+FP}, \tag{20}$$

$$ARMSE = \sqrt{\frac{1}{N}\sum_{n=1}^{N}(b_e - b_g)^2}, \tag{21}$$

$$RRMSE = \sqrt{\frac{1}{N}\sum_{n=1}^{N}(\frac{b_e - b_g}{b_g})^2}, \tag{22}$$

where $b_e$ and $b_g$ denote the estimated and ground truth stenosis ratios, respectively, and $N$ is the number of true positive samples.

Following the guidelines of the Society of Cardiovascular Computed Tomography (SCCT) [39], detected stenosis severity is classified into four categories: minimal (1%–24%), mild (25%–49%), moderate (50%–69%), and severe (70%–100%). To improve clinical relevance, additional constraints are imposed during evaluation. Segments with maximum diameter <1.8 mm or centerline length <20 pixels are excluded as clinically irrelevant. Moreover, if no inflection point with a second derivative of -2 is found, the segment is presumed non-stenotic. If no point with a second derivative of +2 exists, the global maximum diameter in the segment is taken as the reference diameter $d_{ref}$. Only segments with stenosis ratios exceeding 10% are considered significant.

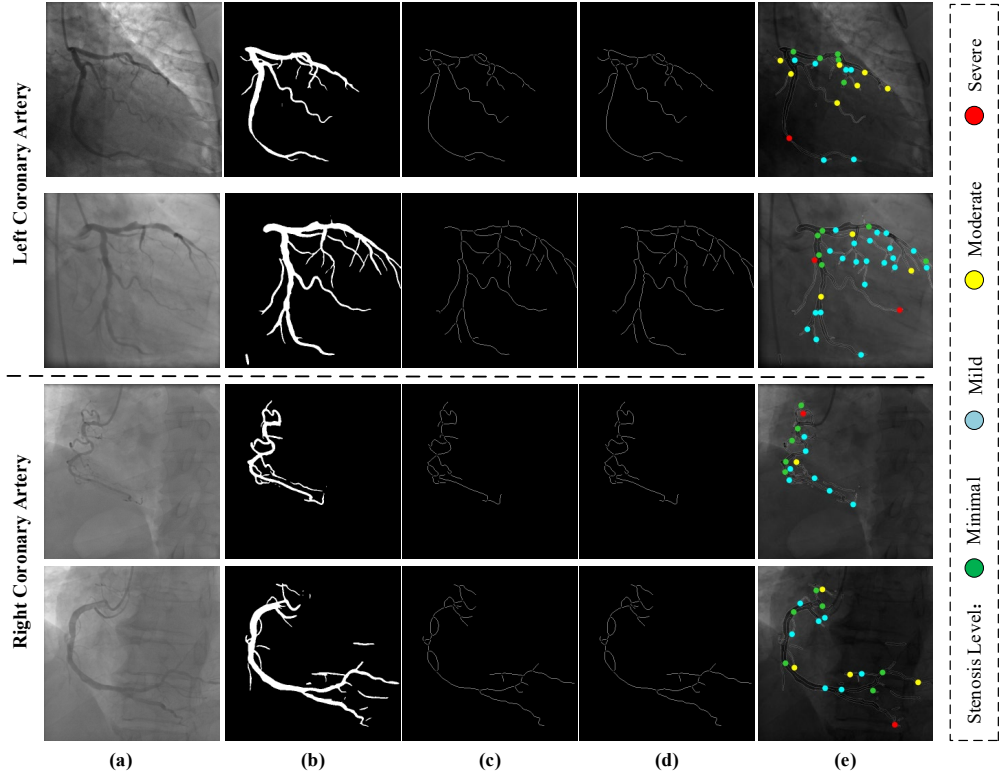

Fig. 10. Visualization of the representative results from the stenosis detection algorithm. (a) Original arterial image; (b) Arterial contours predicted by FAD-Net; (c) Extracted vessel centerlines; (d) Centerline of the selected arterial segment; (e) Detected stenosis points, color-coded by severity: green for minimal, blue for mild, yellow for moderate, and red for severe stenosis.

The hyperparameter detection radius $r$ was empirically set to 10. Fig. 10 illustrates representative results, including centerline extraction, segment selection, and stenosis localization. The contours used were derived from FAD-Net predictions. Quantitatively, the algorithm identified 1,832 true positive points, 1,173 false negatives, and 1,061 false positives across 132 test images. Table 4 summarizes the comparative performance across various stenosis severity levels using contours from different segmentation models.

Table 4. Quantitative evaluation of stenosis detection based on different segmentation models. This table summarizes the performance of arterial segmentation models in detecting stenosis of varying severity levels—minimal (1%-24%), mild (25%--49%), moderate (50%-69%), and severe (70%-100%). Metrics include TPR, PPV, ARMSE, and RRMSE.

| Models | Total Stenosis Points | Detected | Detection Ratio | Percent MSE | TPR↑ | PPV↑ | ARMSE↓ | RRMSE↓ |
|---|---|---|---|---|---|---|---|---|
| | | | | | All / Minimal / Mild / Moderate / Severe | | | |
| U-Net | 3993 | 1525 | 0.3819 | 0.1067 | 0.50 / 0.59 / 0.52 / 0.35 / 0.21 | 0.60 / 0.65 / **0.63** / 0.47 / 0.21 | 0.15 / 0.16 / 0.13 / **0.17** / 0.24 | 0.29 / 0.19 / 0.20 / **0.40** / 2.02 |
| U-Net++ | 4062 | 1625 | 0.4000 | 0.1018 | 0.54 / 0.62 / 0.55 / 0.39 / 0.25 | 0.60 / 0.66 / 0.61 / 0.50 / 0.23 | 0.15 / 0.16 / 0.14 / **0.17** / **0.20** | **0.26** / 0.19 / 0.22 / 0.41 / **1.00** |
| DSCNet | 5942 | 1015 | 0.1708 | 0.1133 | 0.33 / 0.31 / 0.35 / 0.34 / 0.33 | 0.25 / 0.19 / 0.28 / 0.32 / 0.15 | 0.19 / **0.12** / 0.15 / 0.30 / 0.49 | 0.49 / **0.14** / 0.23 / 0.74 / 2.40 |
| Swin-Unet | 3738 | 529 | 0.1415 | 0.1427 | 0.18 / 0.25 / 0.18 / 0.06 / 0.04 | 0.41 / 0.58 / 0.44 / 0.16 / 0.03 | 0.24 / 0.27 / 0.22 / 0.23 / 0.28 | 0.35 / 0.32 / 0.32 / 0.52 / **1.00** |
| FAD-Net | 4066 | 1832 | **0.4505** | **0.0885** | **0.61 / 0.64 / 0.63 / 0.54 / 0.34** | **0.63 / 0.67 / 0.63 / 0.60 / 0.40** | **0.13** / 0.13 / **0.12** / 0.18 / 0.22 | 0.27 / 0.15 / **0.18** / 0.43 / 1.35 |

Results demonstrate that FAD-Net consistently outperforms baseline methods. It achieves an ARMSE of 0.13, an RRMSE of 0.27, a TPR of 0.61, and a PPV of 0.63—showing superior accuracy in both magnitude estimation and classification. Notably, in the severe stenosis category, FAD-Net maintains the lowest ARMSE (0.25) and RRMSE (0.42), while also preserving high sensitivity (TPR = 0.63) and precision (PPV = 0.60). In contrast, methods such as DSCNet and Swim-Unet show significant degradation in detection accuracy, especially in moderate-to-severe categories. Although U-Net++ performs relatively well, its higher RRMSE (up to 0.56 for severe stenosis) highlights its limitations in high-risk lesion recognition. These results affirm the efficacy of the frequency-domain attention and multi-scale LFDM modules in FAD-Net, which significantly contribute to its robustness and clinical potential.

In conclusion, the proposed stenosis detection algorithm, when combined with FAD-Net-generated contours, demonstrates excellent clinical applicability, especially in identifying moderate-to-severe vascular narrowing. It strikes a strong balance between recall and precision, making it a promising candidate for early-stage cardiovascular risk assessment and diagnostic support.

## 5. Discussion

### 5.1. Analysis of Artery Segmentation Performance

The proposed FAD-Net demonstrates notable advantages in coronary artery segmentation from invasive coronary angiography (ICA) images. As observed in Table 2, it achieves achieves the highest DICE score (87.17%), sensitivity (86.34%), and specificity (99.49%), with a significantly reduced HD95 (8.04 mm) and ASSD (1.46 mm) compared to all baselines. This confirms that FAD-Net preserves contour fidelity while minimizing boundary deviation.

The difference map visualization (Fig. 11) reveals that FAD-Net produces substantially fewer false negatives (in red) and false positives (in blue) than other models, underscoring its superior performance in boundary delineation and maintaining continuity, especially in main coronary branches. These improvements can be attributed to the complementary strengths of the MLSA and LFDM modules. Specifically, the MLSA module employs a frequency-domain self-attention mechanism by applying Fourier transforms to multi-scale features. It captures spectral dependencies across low- and high-frequency bands, enhancing the model's responsiveness to vessel-specific structural patterns while suppressing noise and irrelevant activations. This design significantly improves the continuity of thin or low-contrast vessels and reduces segmentation discontinuities by selectively reinforcing high-frequency components associated with fine vascular branches. In parallel, the LFDM module utilizes discrete wavelet transform (DWT) to decompose features into low-frequency (global topology) and high-frequency (edges and textures) components. It then applies progressive diffusion to reinforce the structural integrity of the low-frequency stream, while refining high-frequency details—such as vessel boundaries and stenotic lesions—through cascaded depthwise separable convolutions. These components are then fused via inverse wavelet reconstruction, enabling precise boundary localization and shape preservation. This dual-path refinement strategy contributes directly to improved segmentation accuracy, as evidenced by the significant HD95 reduction (e.g., 1.33 mm vs. 9.36 mm for U-Net++).

As depicted in Fig. 11, FAD-Net achieves better segmentation consistency at vascular bifurcations, which aligns more closely with the ground truth. Since clinical stenosis detection typically focuses on major coronary arteries rather than capillaries, FAD-Net's anatomically continuous contours offer a more reliable basis for subsequent diagnostic tasks.

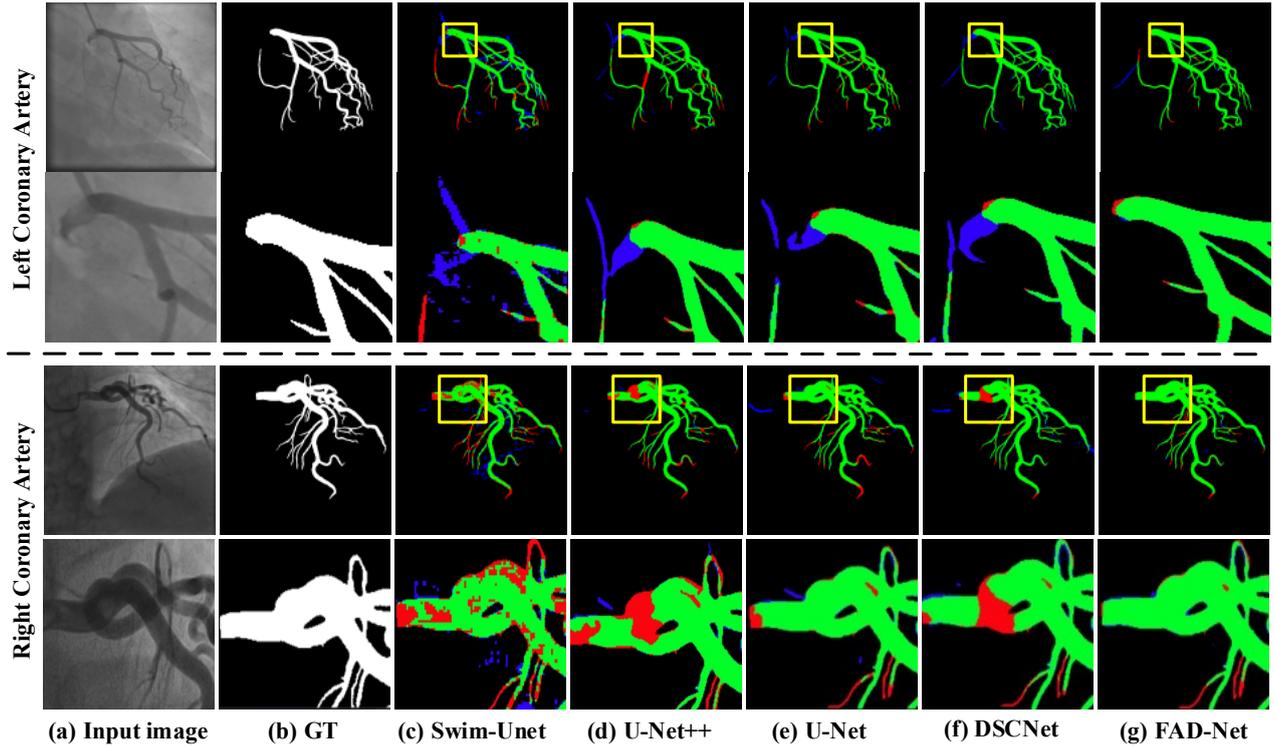

Fig. 11. Visual comparison of segmentation results using different models. (a) Input image; (b) Ground truth; (c-g) Difference maps for Swin-Unet, U-Net++, U-Net, DSCNet, and FAD-Net. Zoomed-in regions highlight model differences. Green, red, and blue regions represent true positives, false negatives, and false positives, respectively.

## 5.2. Analysis of Stenosis Detection Performance

To evaluate the effectiveness of FAD-Net in clinical scenarios, we comprehensively analyze both segmentation quality and stenosis detection accuracy, leveraging visualization (Fig. 12) and quantitative comparison (Table 4). As shown in Fig. 12, FAD-Net demonstrates accurate stenosis localization and severity grading across various coronary arteries. The comparison between ground-truth contours and FAD-Net predictions reveals strong topological alignment, especially in complex bifurcations and distal vessels. Evaluation overlays in subfigure (e) highlight a high density of matched stenosis points (white circles with dark cyan edges) and a notably low number of false negatives (blue) and false positives (red). These errors primarily occur in regions with low vessel contrast or overlapping anatomical structures. Crucially, the model consistently identifies minimal, mild, moderate, and severe stenosis levels, validating its clinical grading capability.

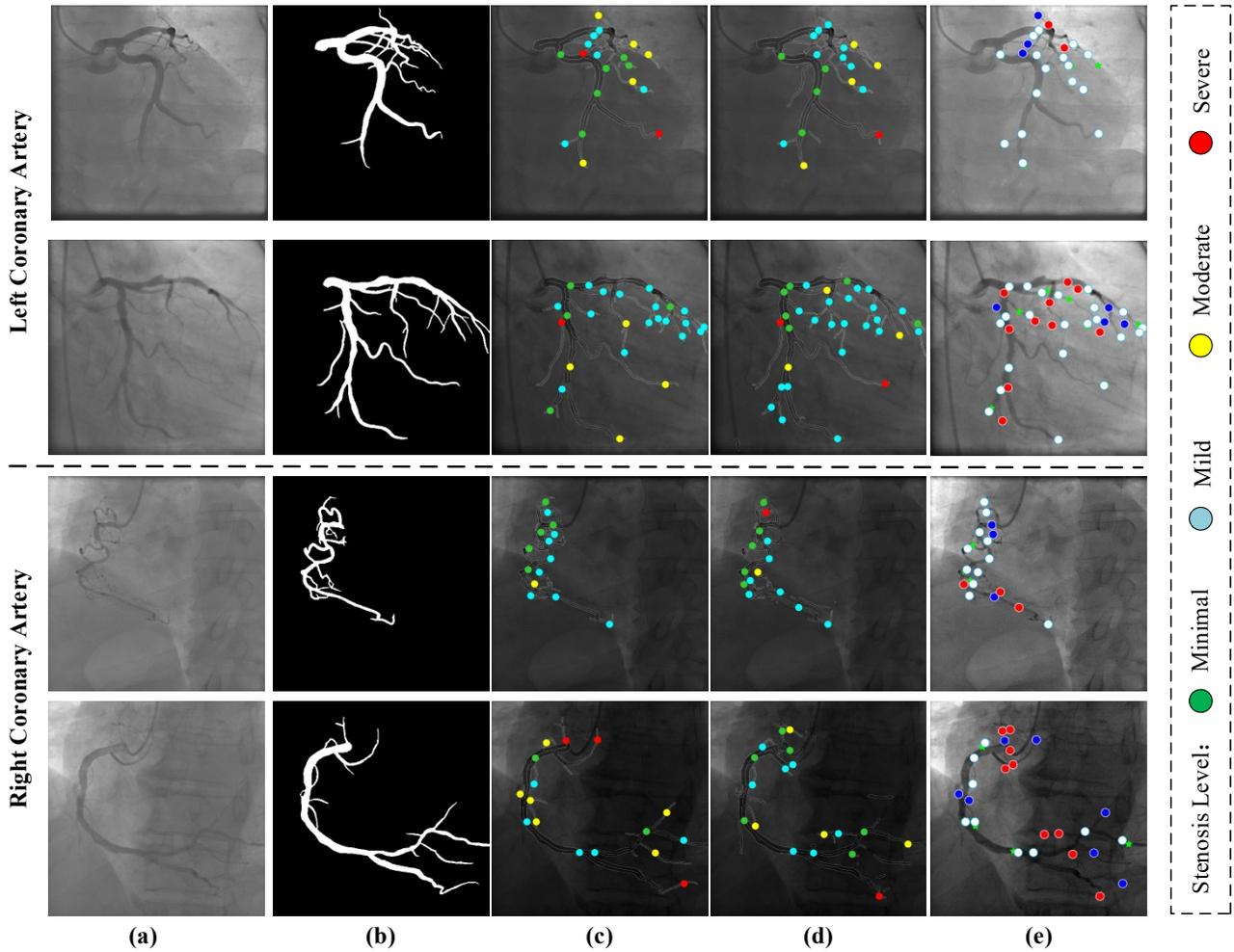

Fig. 12. Stenosis detection results. (a) Original image; (b) Ground-truth contours; (c) Detection using ground-truth contours; (d) Detection using FAD-Net contours; (e) Evaluation. White circles with dark cyan edges: matched points; blue: false negatives; red: false positives. Green, blue, yellow, and red represent minimal, mild, moderate, and severe stenoses, respectively.

Quantitatively, the stenosis detection metrics in Table 4 reinforce FAD-Net's superiority in clinical utility. It achieves the highest overall detection ratio (45.05%) and lowest percent MSE (8.85%). More importantly, FAD-Net shows consistently high true positive rates (TPR) and positive predictive values (PPV) across all severity levels. Specifically, it surpasses all other models in detecting moderate (TPR=0.54) and severe stenoses (TPR=0.34)—which are clinically critical—while maintaining balanced ARMSE (0.13–0.22) and acceptable RRMSE across all levels. This indicates that FAD-Net not only detects more stenosis points but also maintains a high degree of grading accuracy, particularly in moderate-to-severe lesions, which are pivotal for clinical intervention decisions such as stent placement or bypass.

In summary, the grading-aware stenosis detection performance of FAD-Net demonstrate its strong potential as a reliable clinical decision support tool. Its ability to maintain high TPR in severe stenosis and accurately distinguish stenosis grades significantly outperforms existing methods, as supported by both visual and quantitative evidence.

*5.3. Clinical Overview and Applications*

Coronary artery disease (CAD) remains a leading cause of morbidity and mortality worldwide, underscoring the need for accurate diagnosis and timely intervention to improve patient outcomes. Invasive coronary angiography (ICA) is widely regarded as the gold standard for assessing coronary anatomy. However, the conventional manual interpretation of ICA images is inherently subjective and susceptible to inter-observer variability, which can compromise the consistency and reliability of stenosis evaluation.

FAD-Net enhances ICA image segmentation by leveraging frequency-domain processing techniques. Specifically, the MLSA module partitions feature maps into localized regions and applies Fourier transforms to simultaneously amplify high-frequency details and low-frequency structures. This dual enhancement preserves intricate vascular textures while maintaining global structural coherence. In parallel, the LFDM module employs wavelet transforms to enable deep fusion of local and global information, facilitating more precise vascular boundary delineation. Together, these modules significantly improve segmentation accuracy, establishing a more dependable foundation for clinical stenosis assessment.

Clinically, FAD-Net offers several key advantages. It enables automated stenosis detection by producing accurate, reproducible vessel segmentations, thereby supporting objective quantification of diseased regions. In complex scenarios—such as diffuse stenosis, tortuous vessels, or multi-phase ICA acquisitions—FAD-Net demonstrates robust performance, effectively mitigating the limitations of traditional visual analysis. Its frequency-aware architecture also improves resilience to imaging noise and low-contrast conditions, ensuring consistent interpretation across different observers. Furthermore, FAD-Net is readily integrable into clinical CAD software platforms, offering real-time decision support during diagnosis or intervention. Notably, by extending stenosis evaluation from traditional diameter-based measurements to more comprehensive area-based assessments, it provides greater sensitivity to eccentric or irregular lesions. When combined with high-resolution imaging or 3D reconstruction, this approach holds substantial promise for enhancing the clinical evaluation of CAD [40].

*5.4. Limitations*

Despite its encouraging performance, several limitations of FAD-Net should be acknowledged. First, the training dataset remains relatively limited in size and diversity. Manual annotation of ICA images is resource-intensive and time-consuming, which constrains the availability of large-scale, high-quality labeled data and may limit the model's generalizability across broader clinical scenarios. Second, the current stenosis detection framework is based on static, single-view ICA frames. In clinical practice, however, multi-view and dynamic imaging sequences are often essential for capturing complex spatial relationships—especially in the presence of vessel overlap or foreshortening. The reliance on static views may thus reduce detection accuracy in such cases. Third, ICA inherently provides two-dimensional projections of three-dimensional coronary structures. This dimensionality reduction can introduce geometric distortions that affect the fidelity of quantitative analysis, potentially impacting the reliability of stenosis grading [41].

Future work will focus on addressing these limitations by expanding the dataset to include more diverse cases, incorporating temporal and multi-angle ICA sequences, and exploring 3D reconstruction techniques. These efforts aim to further enhance the robustness and clinical applicability of FAD-Net.

## 6. Conclusion

In this study, we proposed FAD-Net, a novel deep learning framework that integrates Multi-Level Self-Attention (MLSA) module and Low-Frequency Diffusion Module (LFDM) within a U-shaped architecture for coronary artery segmentation in invasive coronary angiography (ICA) images. By jointly exploiting global and local frequency-domain features, FAD-Net effectively overcomes key challenges such as low contrast, imaging noise, and complex vascular morphology. Extensive experiments on a dataset comprising 616 multi-view ICA images from 99 patients demonstrate that FAD-Net achieves superior segmentation performance compared to existing state-of-the-art models. In addition, the proposed stenosis detection pipeline—which includes centerline extraction, keypoint localization, and quantitative assessment of vascular narrowing—further enhances lesion characterization and supports objective, reproducible clinical decision-making.

Overall, FAD-Net exhibits strong potential for clinical integration in computer-aided diagnosis and treatment planning for coronary artery disease. By incorporating frequency-domain processing into the segmentation paradigm, our approach offers a robust, accurate, and interpretable solution for cardiovascular image analysis, and lays a foundation for future applications in vessel segmentation across various imaging modalities.

**Acknowledgment**

Nan Mu was supported by the Natural Science Foundation of Sichuan Province (2025ZNSFSC1477) and the National Natural Science Foundation of China (62006165).